\documentstyle[preprint,pre,aps,eqsecnum]{revtex}
\tighten    

\begin{document}
\draft

\preprint{Submitted to Physical Review E}

\title{Chaos for Liouville probability densities}

\author{R\"udiger Schack\cite{MileEnd} and Carlton M. Caves}
\address{Center for Advanced Studies,
Department of Physics and Astronomy,\\
University of New Mexico, Albuquerque, NM~87131-1156}
\date{\today}
\maketitle

\begin{abstract}
Using the method of symbolic dynamics, we show that a large class of
classical chaotic maps exhibit {\it exponential hypersensitivity to
perturbation}, i.e., a rapid increase with time of the information
needed to describe the perturbed time evolution of the Liouville
density, the information attaining values that are exponentially
larger than the entropy increase that results from averaging over the
perturbation.  The exponential rate of growth of the ratio of information
to entropy is given by the Kolmogorov-Sinai entropy of the map.  These
findings generalize and extend results obtained for the baker's map
[R.~Schack and C.~M. Caves, Phys.\ Rev.\ Lett.\ {\bf 69}, 3413 (1992)].
\end{abstract}


\section{Introduction}

Chaos in Hamiltonian systems is usually defined in terms of trajectories
of phase-space points. The Lyapunov exponent describes how initially
close trajectories diverge exponentially\cite{Lichtenberg1983}.  The
Kolmogorov-Sinai (KS) entropy measures the rate at which information
about the initial phase-space point must be supplied in order to
predict the coarse-grained behavior of a trajectory at a later time
\cite{Alekseev1981,Ford1983}.

Signatures of chaos are less obvious if attention is shifted from the
time evolution of phase-space points to the time evolution of probability
densities, governed by the Liouville equation. If the distance between
two densities is defined in terms of an overlap integral, there is no
exponential divergence of initially close densities since the overlap
integral is constant in time (``Koopman's theorem''
\cite{Koopman1931,Peres1993a}). Furthermore, as a direct consequence of
Koopman's theorem, if one is given the Hamiltonian and the initial density
to a certain accuracy, then no additional information is needed to predict
the density at all later times $t$ to the same accuracy, except for a
negligible amount of information needed to specify the time $t$
\cite{Zurek1989b,Caves1990c}. This means that the popular
information-theoretic interpretation \cite{Ford1983} of chaos via
the KS entropy does not apply to Liouville densities.

In this paper we show that there is an information-theoretic way to
characterize chaos for Liouville densities in systems with a positive
KS entropy.  In particular, we show that a large class of Hamiltonian
systems with positive KS entropy display an exponential hypersensitivity
to perturbation.  We have investigated hypersensitivity
to perturbation previously \cite{Caves1993b,Caves1993a,Schack1992a,%
Schack1993e,Schack1994b}, both for classical and quantum systems, and
have characterized it is as a rapid increase with time of the information
needed to describe the perturbed time evolution of the system state
(Liouville density for classical systems, state vector for quantum
systems), the information attaining values much larger than the entropy
increase that results from averaging over the perturbation.

Here we formulate the concept of hypersensitivity to perturbation more
precisely.  We consider the amount of information needed to keep
track of the perturbed time evolution to a level of accuracy that
keeps the increase of system entropy below a certain ``tolerable'' level.
This information should be compared to the entropy reduction it buys,
i.e., to the difference between the entropy increase that results from
averaging over the perturbation and the tolerable entropy increase.  We
characterize hypersensitivity to perturbation in terms of the ratio
of information to entropy reduction.  A system displays
{\it hypersensitivity to perturbation\/} if the ratio grows rapidly with
time, becoming much larger than unity, for almost all values of the
tolerable entropy; a system displays {\it exponential hypersensitivity
to perturbation\/} if the ratio grows exponentially.  We show that
a large class of Hamiltonian systems with positive KS entropy display
exponential hypersensitivity to perturbation, with the exponential
growth rate given by the KS entropy.  This result establishes a direct
connection between measures of chaos based on trajectories and our
information-theoretic characterization for Liouville densities.

There are at least two important motivations for investigating signatures
of chaos in Liouville densities. One motivation comes from the tricky
question of how to characterize quantum chaos.  In quantum mechanics,
trajectories of state vectors show no sensitivity to initial conditions
because the Schr\"odinger equation is linear and preserves the inner
product. This argument does not prove, however, that there is no chaos
in quantum mechanics \cite{Berry1992}, because the Liouville equation,
like the Schr\"odinger equation, is linear and preserves the overlap
between densities, yet any chaotic classical Hamiltonian system can
be described by a Liouville equation. Furthermore, the classical analog
of a quantum state vector is not a point in classical phase space, but
a Liouville density\cite{Peres1993a,Caves1993a}. In contrast to the
above-mentioned characterizations of classical chaos in terms of phase-space
trajectories, a characterization of classical chaos in terms of Liouville
densities can be expected to have a straightforward generalization to
quantum systems \cite{Caves1993b}. We have indeed found that hypersensitivity
to perturbation is present in quantum systems \cite{Schack1993e,Schack1994b}.

The other main motivation for studying chaos in Liouville densities lies
in the central role Liouville densities play in statistical mechanics.
The connection of the present work with statistical mechanics is outlined
in Sec.~\ref{secstat}.  In Sec.~\ref{sechyp} we give a precise definition
of hypersensitivity to perturbation.  Section~\ref{secsym} reviews the
method of symbolic dynamics.  In Sec.~\ref{secper}, the heart of the
paper, we apply the method of symbolic dynamics to prove that a large
class of perturbed chaotic systems display exponential hypersensitivity
to perturbation.  In Sec.~\ref{discussion} we distill the essence of
the symbolic-dynamics analysis to develop a simple, heuristic picture of
hypersensitivity to perturbation, which explains why chaotic systems
exhibit exponential hypersensitivity to perturbation and regular, or
integrable systems do not.  A reader not interested in the details of
the symbolic dynamics might profitably skip Secs.~\ref{secsym} and
\ref{secper} and proceed directly to Sec.~\ref{discussion}.

\section{Connection with statistical mechanics}   \label{secstat}

In statistical mechanics the exact point a system occupies in phase space
typically is not known. The predictions of classical statistical
mechanics are derived from a Liouville probability density $\rho(x)$ on
phase space, which describes incomplete knowledge of the system's
phase-space point $x$ and which is the mathematical representation of a
system state. The entropy (in bits) of a system state $\rho(x)$, also
called the Gibbs entropy or fine-grained entropy, is defined as
\begin{equation}
H=-\int d\Gamma(x)\,\rho(x)\log_2 [\rho(x)]\;,
\label{Gibbsent}
\end{equation}
where $\Gamma(x)$ is the standard phase-space measure. (The use of
base-2 logarithms here and throughout this paper means that entropy
and information are measured in bits.) Since the Gibbs entropy is
formally identical to Shannon's \cite{Shannon1948} statistical measure
of information, entropy can be interpreted as the amount of information
missing toward a complete specification of the system.  The classical
entropy is defined up to an arbitrary additive constant, reflecting the
fact that an infinite amount of information is needed to give the exact
location of a point in phase space.

As a consequence of Liouville's theorem, the entropy remains constant
under Hamiltonian time evolution. We adopt here the Bayesian, or
information-theoretic, approach to statistical mechanics
\cite{Jaynes1957b,Jaynes1957a,Jaynes1983}, according to which
the constancy of the Gibbs entropy is an expression of the fact
that no information about the initial Liouville density is lost under
Hamiltonian time evolution.

The Bayesian approach to statistical mechanics is connected with
thermodynamics in the following way: Assume there is a heat reservoir
at temperature $T$, with which all energy in the form of heat must
ultimately be exchanged, possibly by using intermediate steps such as
storage at some other temperature; then each bit of missing information
about the system state reduces by the amount $k_BT\ln2$ the energy that
can be extracted from the system in the form of useful work. The Bayesian
approach can thus be summarized in two statements: (i)~entropy is missing
information---a mathematical statement; (ii)~each bit of missing
information costs $k_BT\ln2$ of useful work---this is the physics.

Since entropy is a measure of missing information, entropy increases if
information about the system is lost. There are two main mechanisms
leading to information loss (as noted above, Hamiltonian time evolution
is {\it not\/} such a mechanism): deliberate discarding of information
and loss of information through interaction with an incompletely known
environment.

Deliberate discarding of information was used by Jaynes
\cite{Jaynes1957b,Jaynes1957a,Jaynes1983} to derive traditional
thermodynamics. Jaynes showed how equilibrium thermodynamics follows
effortlessly from the Liouville equation if the only information retained
is the values of the macroscopic variables defining a thermodynamic state.
In Jaynes's approach, irrelevant information is discarded by means of the
principle of maximum entropy. Another example is the derivation of the
Boltzmann equation\cite{Spohn1991}; here information about correlations
between particles is discarded as irrelevant.

In contrast to these examples where information is discarded deliberately,
an actual loss of information can occur in a system that, rather than being
perfectly isolated, interacts with an incompletely known environment. The
interaction with the environment leads to a perturbed time evolution of the
system. Predictions for the system alone are made by tracing out the
environment---i.e., by averaging over the perturbations---which generally
leads to an entropy increase.  This approach was pioneered by Borel
\cite{Borel1914,Borel1925}. The entropy increase of the system due to the
interaction with the environment is a result of the environment's being in
an at least partially unknown state. If suitable information about the
environment is obtained, the increase in system entropy can be reduced
or, if sufficient information is obtained, prevented entirely.  Averaging
over the perturbing environment is usually justified by arguing that it
is impossible in practice to control the environment.

In this paper we go beyond the pragmatic argument that controlling the
interaction with the environment is impossible in practice. We show how the
information-theoretic approach to statistical mechanics leads naturally to a
quantitative measure of how hard it is to keep the entropy of the system from
increasing by gathering information about the environment. The key to
quantifying the difficulty of controlling the interaction with the environment
is Landauer's principle\cite{Landauer1961,Landauer1988}, which assigns a
thermodynamic cost to information.  According to Landauer's principle, in
the presence of a heat reservoir at temperature $T$, not only does each
bit of {\it missing\/} information have a free-energy cost of $k_BT\ln2$,
but each bit of information that is {\it acquired\/} has the same free-energy
cost of $k_BT\ln2$.  This cost, called the {\it Landauer erasure cost},
is paid when the acquired information is erased.  Acquired information can
be quantified by algorithmic information
\cite{Zurek1989b,Caves1990c,Chaitin1987a,Zurek1989a,Schack1994a};
roughly speaking, the algorithmic information in an observational record
is the length in bits of the shortest record having the same information
content.

The question of how hard it is to reduce the system entropy by controlling
the environment can now be given a quantitative form: ``How big is the
Landauer erasure cost of the information {\it about the environment\/}
which is needed to reduce the increase of system entropy by a certain
amount?'' In the next section we give a mathematical formulation of
this question. The later parts of this paper are devoted to showing that
the answer can be used to characterize chaos.

\section{Hypersensitivity to perturbation}   \label{sechyp}

Consider a classical Hamiltonian system initially described by a Liouville
density $\rho(x,\mbox{$t=0$})$ on phase space. The initial entropy is
\begin{equation}
H_0=-\int d\Gamma(x)\,\rho(x,\mbox{$t=0$})\log_2[\rho(x,\mbox{$t=0$})] \;,
\end{equation}
where $\Gamma(x)$ is the standard phase-space measure. By solving the Liouville
equation, one obtains the density $\rho(x,t)$ at time $t$. According to
Liouville's theorem, the entropy remains unchanged---the information about the
initial density is preserved under Hamiltonian time evolution.

Now assume that the system is coupled to an incompletely known environment
in such a way that the interaction can be described as an energy-conserving,
typically time-dependent perturbation of the system Hamiltonian. The
system's interaction with the environment is thus described by a
stochastic Hamiltonian. We denote the perturbed system state at time
$t$ by $\rho_y(x,t)$ where $y$ labels the particular realization of the
stochastic perturbation or {\it perturbation history}. The possible
perturbation histories $y$ are distributed according to a probability
measure $\gamma(y)$. This description in terms of a stochastic system
Hamiltonian applies when the system is coupled to conserved quantities
of the environment. The values of the conserved environment quantities
label the perturbation histories $y$, and the probability measure
$\gamma(y)$ is the probability measure for the conserved environment
quantities.

For each perturbation history $y$, the entropy of the density
$\rho_y(x,t)$ is equal to the initial entropy $H_0$. Averaging over all
possible perturbation histories leads to an average density
\begin{equation}
\bar\rho(x,t) =\int d\gamma(y)\,\rho_y(x,t)\;,
\end{equation}
with entropy
\begin{equation}
\bar H=-\int
d\Gamma(x)\,\bar\rho(x,t)\log_2[\bar\rho(x,t)] \equiv H_0+\Delta H_{\cal S}\;,
\end{equation}
where $\Delta H_{\cal S}\geq0$ is the entropy increase due to averaging
over the incompletely known environment.  That $\Delta H_{\cal S}\geq0$
follows from the concavity of the entropy: the entropy of an average
distribution is greater than or equal to the average entropy of the
distributions that contribute to the average.

Now assume, in accordance with the discussion of Sec.~\ref{secstat} about
gathering information from the environment, that an arbitrary measurement,
with discrete possible outcomes labeled by integers $b$, is performed on
the environment. The outcome $b$ has conditional probability $p_{b|y}$,
given the perturbation history $y$, and hence has unconditioned probability
\begin{equation}
p_b=\int d\gamma(y)\,p_{b|y}\;.
\end{equation}
The Liouville density for the system state conditional on outcome $b$
we denote by
\begin{equation}
\rho_b(x,t)={1\over p_b}\int d\gamma(y)\,\rho_y(x,t)p_{b|y}
=\int d\gamma(y|b)\,\rho_y(x,t)\;,
\label{rhosubbx}
\end{equation}
where $\gamma(y|b)$ is the probability measure for the perturbation
histories conditional on outcome $b$.  It follows immediately that
\begin{equation}
\sum_b p_b\rho_b(x,t)=\int d\gamma(y)\,\rho_y(x,t)=\bar\rho(x,t) \;.
\label{rhobarb}
\end{equation}

We denote by
\begin{equation}
\Delta H_b=-\int d\Gamma(x)\,\rho_b(x,t)\log_2[\rho_b(x,t)] - H_0 \geq 0
\end{equation}
the change in system entropy conditional on the measurement outcome $b$,
where the inequality follows from applying concavity to Eq.~(\ref{rhosubbx}),
and by
\begin{equation}
\overline{\Delta H} = \sum_b p_b\Delta H_b \leq \Delta H_{\cal S}
\end{equation}
the average conditional entropy change, where the inequality follows
from applying concavity to Eq.~(\ref{rhobarb}).  Finally, we denote by
\begin{equation}
\overline{\Delta I} = -\sum_b p_b \log_2 p_b \label{eqdeli}
\end{equation}
the average information needed to specify the measurement outcome~$b$.
Actually, Eq.~(\ref{eqdeli}) is only a lower bound to the average
algorithmic information needed to specify the measurement outcome $b$,
but it can be shown to be an extremely tight lower
bound\cite{Schack1994a}.  An immediate consequence of the definition
of entropy is that
\begin{equation}
\overline{\Delta I}+\overline{\Delta H}\geq\Delta H_{\cal S}\;,
\label{totalentropy}
\end{equation}
with equality holding if and only if the densities $\rho_b(x,t)$ are
disjoint.

Suppose now that one wants to limit the entropy increase of the system
to a certain {\it tolerable amount\/} $\Delta H_{\rm tol}$.  Then the
minimum amount of information about the perturbing environment needed
to keep the system entropy from increasing by more than
$\Delta H_{\rm tol}$ can be written as
\begin{equation}
\Delta I_{\rm min}\equiv\inf_{\overline{\Delta H}\leq\Delta H_{\rm tol}}
   \overline{\Delta I} \;,
\end{equation}
where the infimum is taken over all possible measurement schemes for
which the average conditional entropy increase does not exceed
$\Delta H_{\rm tol}$.  In other words, $\Delta I_{\rm min}$ is the
information about the environment that it takes to lower the entropy
increase of the system from $\Delta H_{\cal S}$ (the increase due to
averaging over the perturbation) down to $\Delta H_{\rm tol}$;
i.e., $\Delta I_{\rm min}$ is the minimum information about the
environment needed to reduce the system entropy by an amount
$\Delta H_{\cal S}-\Delta H_{\rm tol}$. As a consequence of
Eq.~(\ref{totalentropy}), it is a general theorem---and an expression
of the second law---that
\begin{equation}
\Delta I_{\rm min} \geq \Delta H_{\cal S}-\Delta H_{\rm tol} \;.
\end{equation}
In the presence of a heat reservoir at temperature $T$, the information
$\Delta I_{\rm min}$ has an energy cost $k_BT\ln2\,\Delta I_{\rm min}$
on erasure, which should be compared to the gain in extractable work due
to the observation, $k_BT\ln2\,(\Delta H_{\cal S}-\Delta H_{\rm tol})$.

We are now in a position to define hypersensitivity to perturbation.
We say a system is {\it hypersensitive to perturbation\/} if, for almost
all values of $\Delta H_{\rm tol}$, the information $\Delta I_{\rm min}$
is large compared with the corresponding entropy reduction
$\Delta H_{\cal S}-\Delta H_{\rm tol}$, i.e.,
\begin{equation}
{\Delta I_{\rm min}\over\Delta H_{\cal S}-\Delta H_{\rm tol}}\gg1\;.
\label{eqhyp}
\end{equation}
In terms of energy this definition says that, for a system displaying
hypersensitivity to perturbation, possible gains in system free energy
through observations of the environment are negligible compared to the
Landauer erasure cost of the observational records.

Hypersensitivity to perturbation requires that the inequality~(\ref{eqhyp})
hold for almost all values of $\Delta H_{\rm tol}$.  The
inequality~(\ref{eqhyp}) tends always to hold for sufficiently small
values of $\Delta H_{\rm tol}$.  The reason is that for these small
values of $\Delta H_{\rm tol}$, one is gathering enough information
from the perturbing environment to track a particular system state
whose entropy is nearly equal to the initial system entropy $H_0$.
In other words, one is essentially tracking a particular realization
of the perturbation among all possible realizations.  Thus, for
small values of $\Delta H_{\rm tol}$, the information
$\Delta I_{\rm min}$ is a property of the perturbation, being the
information to specify a particular realization of the perturbation.
The important regime for assessing hypersensitivity to perturbation is
thus where $\Delta H_{\rm tol}$ is near to $\Delta H_{\cal S}$, and it
is in this regime that one can hope that $\Delta I_{\rm min}$ reveals
something about the system dynamics, rather than properties of the
perturbation.

In earlier publications\cite{Caves1993b,Schack1992a,Schack1993e,Schack1994b},
we have conjectured that chaotic Hamiltonian systems, classical or
quantum, show hypersensitivity to perturbation. For classical chaotic
systems, this can be made plausible in the following way.  Under chaotic
time evolution, the Liouville density develops structure on finer and
finer scales.  This highly structured pattern is not itself complex in
the algorithmic sense---it is completely specified by the initial density,
the Hamiltonian, and the elapsed time---but it can be perturbed in an
enormous number of ways \cite{Caves1993b}.  This means that the
unperturbed pattern lies very close to a large number of highly
complex patterns and that the information about the perturbation needed
to specify the perturbed pattern can be very large.  In Sec.~\ref{secper}
we go beyond this heuristic argument and give a proof that a large class
of classical chaotic Hamiltonian systems exhibit an {\it exponential\/}
hypersensitivity to perturbation, in which the ratio~(\ref{eqhyp}) of
information to entropy reduction grows exponentially with time, with
the exponential rate of growth given by the KS entropy of the chaotic
dynamics.  We find that for this class of chaotic systems, the exponential
hypersensitivity to perturbation is to a large extent independent of
the exact nature of the perturbations and, in particular, of the
strength of the perturbations.

In the following sections we limit our investigation to discrete maps.
There are two natural ways in which a Hamiltonian flow
$\phi_t:X\rightarrow X$ on the phase space $X$ induces a discrete map.
For an arbitrary time step $\tau$, a map $f:X\rightarrow X$ is defined
by $fx=\phi_\tau x$ for all $x\in X$. Since $\phi_t\phi_s x=\phi_{t+s}x$
for all times $t$ and $s$ and all $x\in X$, the map $f$ and the flow
$\phi_t$ are closely related by $f^nx=\phi_{n\tau}x$ for all $x\in X$
and all integer $n$. Alternatively, a discrete map can be defined via
a Poincar\'e surface of section\cite{Lichtenberg1983}. The stochastic
perturbation of the flow induces a stochastic perturbation of the map
at each step.

\section{Symbolic dynamics}  \label{secsym}

The basic idea underlying the method of symbolic dynamics is to
simplify the analysis of dynamical systems by representing points in
phase space by symbolic sequences. Parts of the following discussion
closely follow \cite{Alekseev1981}.

A {\it discrete abstract dynamical system\/} $(M,\mu,f)$ consists of a
measurable space $M$ with a normalized measure $\mu$ and a measure-preserving
automorphism $f$ on $M$, i.e., $\mu(M)=1$ and $\mu(f A)=\mu(A)$ for all
measurable $A$\cite{Arnold1968,Ornstein1991}.  A {\it measurable
partition\/} $\cal E$ of $M$ is defined as a collection
${\cal E} =\{E_1,\ldots,E_m\}$ of measurable sets such that
\begin{equation}
\bigcup_{i=1}^m E_i=M\;{\rm~and~}\;\sum_{i=1}^m\mu(E_i)=1\;.
\end{equation}
Consider an $m$-letter alphabet ${\cal L}=\{1,\ldots,m\}$ where each
letter corresponds to one of the $m$ sets in the partition $\cal E$. We
denote by $\omega =\cdots\omega_{-1}\omega_0\omega_1\omega_2\cdots$ a
bi-infinite sequence of letters $\omega_n\in\cal L$ and by $\Sigma$ the
set of all such {\it symbolic sequences}.

For each $x\in M$ we define the set $\Sigma_x\subseteq\Sigma$ as follows:
\begin{equation}
\Sigma_x\equiv
\biggl\{\omega\;\biggl|\;
x\in\bigcap_{n=-\infty}^\infty f^{-n}E_{\omega_n}\biggr\}\;.
\label{sigmax}
\end{equation}
Equivalently, one can say that
$\omega\in\Sigma_x\Longleftrightarrow f^nx\in E_{\omega_n}$ for all $n$.
The set
\begin{equation}
\Sigma_{\cal E} =\bigcup_{x\in M}\Sigma_x\subseteq\Sigma
\end{equation}
of all symbolic sequences corresponding to at least one point in $M$ is
called the set of {\it admissible sequences}. The partition $\cal E$ is
called a {\it generating partition\/} if for each $\omega\in\Sigma_{\cal E}$
the intersection
\begin{equation}
\bigcap_{n=-\infty}^\infty f^{-n}E_{\omega_n}
\label{eqinft}
\end{equation}
consists of only one point, i.e., if each admissible symbolic sequence
defines a unique point in $M$.  In general, even for generating partitions,
$\Sigma_x$ may have more than one element, which means that a point
$x\in M$ may be represented by several symbolic sequences $\omega\in\Sigma_x$.
For a generating partition, the picture one should have is that the
set $\Sigma_{\cal E}$ of all admissible sequences is the union of
disjoint subsets $\Sigma_x$, which may have more than one member.

Let us further define {\it symbolic words} as finite symbolic sequences
$\omega_{n_1}\cdots\omega_{n_2}$ where $n_1\leq n_2$. In analogy with
Eq.~(\ref{eqinft}), we define the set of points corresponding to the
symbolic word $\omega_{n_1}\cdots\omega_{n_2}$ by
\begin{equation}
E_{\omega_{n_1}\cdots\omega_{n_2}}
     =\bigcap_{n=n_1}^{n_2}f^{-n}E_{\omega_n}\;.
\label{Eword}
\end{equation}
We denote by $\Sigma^{(n_1,n_2)}$ the set of all symbolic words
$\omega_{n_1}\cdots\omega_{n_2}$.  The symbolic word
$\omega_{n_1}\cdots\omega_{n_2}$ is {\it admissible\/} if
$E_{\omega_{n_1}\cdots\omega_{n_2}}$ contains at least one point; we
denote by $\Sigma_{\cal E}^{(n_1,n_2)}$ the set of admissible symbolic
words $\omega_{n_1}\cdots\omega_{n_2}$.  The $N$th refinement
${\cal E}^N$ of the partition ${\cal E}$, defined by
\begin{equation}
{\cal E}^N =\{E_{\omega_0\cdots\omega_{N-1}}\;|\;
  \omega_0\cdots\omega_{N-1}\;{\rm admissible}\}\;,
\end{equation}
is also a measurable partition. If $\cal E$ is a generating partition,
then all refinements of $\cal E$ are also generating partitions.
Furthermore, if $\cal E$ is generating, then the sigma algebra generated
by all refinements of $\cal E$ coincides with the sigma algebra of all
measurable subsets of $M$
\cite{Kolmogoroff1958,Kolmogoroff1959,Crutchfield1983}. The measure $\mu$
induces a measure on the sigma algebra generated by the set of all symbolic
words via
\begin{equation}
\mu(\omega_{n_1}\cdots\omega_{n_2})
          =\mu(E_{\omega_{n_1}\cdots\omega_{n_2}}) \;.
\end{equation}
Let us also define a conditional measure
\begin{equation}
\mu(\omega_{n_1}\cdots\omega_{n_2}|\omega_{n_2+1}\omega_{n_2+2}\cdots)
    = \lim_{n\rightarrow\infty}
{\mu(\omega_{n_1}\cdots\omega_{n_2+n}) \over
              \mu(\omega_{n_2+1}\cdots\omega_{n_2+n})}
\end{equation}
whenever the limit on the right-hand side exists, which is the case for
$K$ systems (see below).

The {\it entropy\/} $H({\cal E}^N)$ of the refinement ${\cal E}^N$ is
defined by
\begin{equation}
H({\cal E}^N) = -\sum_{E_{\omega_0\cdots\omega_{N-1}}\in{\cal E}^N}
  \mu(E_{\omega_0\cdots\omega_{N-1}})
  \log_2\mu(E_{\omega_0\cdots\omega_{N-1}}) \;.
\end{equation}
The {\it metric entropy\/} or {\it Kolmogorov-Sinai (KS) entropy\/} of
the map $f$ is defined as
\begin{equation}
h_{\mu}(f)=\sup_{\cal E} h_{\mu}(f|{\cal E})\;,
\end{equation}
where the supremum is taken over all measurable partitions $\cal E$
and where
\begin{equation}
h_{\mu}(f|{\cal E})
    = \lim_{N\rightarrow\infty} \frac{H({\cal E}^N)}{N} \;.
\end{equation}
If $\cal E$ is generating, then $h_\mu(f)=h_\mu(f|{\cal E})$
\cite{Kolmogoroff1958,Kolmogoroff1959,Crutchfield1983}.  Systems with a
positive KS entropy are called {\it $K$ systems}.  Despite its name,
the KS entropy is quite different from the Gibbs entropy, for two reasons:
(i)~$H({\cal E}^N)$ has nothing directly to do with probabilities on
the phase space $M$, but is the Shannon information of the ensemble
of sets in ${\cal E}^N$, when the probability of each set is given
by its measure; (ii)~$h_\mu(f|{\cal E})$ is not an entropy at all,
but rather is the asymptotic {\it rate of increase\/} of $H({\cal E}^N)$.

A dynamical system is called {\it ergodic\/} if time averages equal
ensemble averages, i.e., if
\begin{equation}
\lim_{N\rightarrow\infty}\frac{1}{N}\sum_{n=0}^{N-1}\phi(f^n x) =
  \int_M d\mu\,\phi \;\;\;\mbox{for almost all $x\in M$,}
\label{eqerg}
\end{equation}
for any $\mu$-integrable function $\phi$\cite{Arnold1968}.  All $K$
systems are ergodic\cite{Arnold1968}.

The map $f$ induces a particularly simple so-called {\it shift map\/}
$\sigma:\Sigma\rightarrow\Sigma$ on the set of symbolic sequences. The shift
map is defined as
\begin{equation}
(\sigma\omega)_n=\omega_{n+1}
\mbox{~for all $n$;}
\end{equation}
i.e., $\sigma$ shifts the entire symbolic sequence to the left.  The shift
map can be extended to a map
$\sigma:\Sigma^{(n_1,n_2)} \rightarrow \Sigma^{(n_1-1,n_2-1)}$ that acts on
symbolic words $\omega_{n_1}\cdots\omega_{n_2}\in\Sigma^{(n_1,n_2)}$ via
\begin{equation}
[\sigma(\omega_{n_1}\cdots\omega_{n_2})]_n =\omega_{n+1}
\mbox{~for $n_1-1<n<n_2-1$.}
\end{equation}
The set of admissible sequences is invariant under the shift map, i.e.,
\begin{equation}
\sigma(\Sigma_{\cal E})=\Sigma_{\cal E} \;.
\label{eqinv}
\end{equation}
Furthermore, for a generating partition  $\cal E$, the map
$\pi:\Sigma_{\cal E} \rightarrow M$ defined by
\begin{equation}
\pi(\omega)=\bigcap_{n=-\infty}^\infty f^{-n}E_{\omega_n}
\end{equation}
[i.e., $\pi(\omega)=x\Longleftrightarrow\omega\in\Sigma_x$]
is single-valued and continuous\cite{Alekseev1981}. If the sets $E_i$
forming the partition $\cal E$ are not mutually exclusive, then the
map $\pi$ is not one-to-one. The overlap between different sets $E_i$,
however, is of measure zero.  The relation between $f$ and $\sigma$
can be summarized in the following commutation diagram:
\begin{equation}
\begin{array}{rcl}
\Sigma_{\cal E}&\stackrel{\sigma}{\longrightarrow}&\Sigma_{\cal E}\\
{\scriptstyle\pi}\downarrow && \downarrow{\scriptstyle\pi} \\
M & \stackrel{f}{\longrightarrow} & M
\end{array}  \;.
\label{eqdiagram}
\end{equation}
The action of $f$ on measurable subsets of $M$ is faithfully represented
by the action of $\sigma$ on measurable sets of symbolic sequences.
In the following section, we use this representation to study
hypersensitivity to perturbation for $K$ systems.

For the remainder of this section, we assume that $f$ is a $K$ system
with KS entropy $h$ and that $\cal E$ is a generating partition. Since
the set $\Sigma_{\cal E}$ of admissible symbolic sequences is invariant
under the action of the shift map $\sigma$ according to Eq.~(\ref{eqinv}),
$\Sigma_{\cal E}$ is a stationary source in the language of information
theory \cite{Welsh1988}.  Moreover, by choosing the function $\phi$ in
Eq.~(\ref{eqerg}), for an arbitrary symbolic word
$\tilde\omega=\omega_{n_1}\ldots\omega_{n_2}$, to be the indicator
function of the set $E_{\tilde\omega}$ [see Eq.~(\ref{Eword})]
corresponding to $\tilde\omega$,
i.e.,
\begin{equation}
\phi_{\tilde\omega}(x)= \left\{  \begin{array}{ll}
1\;, & \mbox{if $x\in E_{\tilde\omega}$,} \\
0\;, & \mbox{otherwise,}   \end{array}  \right.
\end{equation}
one sees that $\Sigma_{\cal E}$ is an ergodic source since $f$ is ergodic.

According to the Shannon-McMillan theorem, stationary ergodic sources have
the {\it asymptotic equipartition property\/} \cite{Welsh1988}. This means
crudely that for sufficiently large $n$ and arbitrary $n_1$, the set
$\Sigma_{\cal E}^{(n_1,n_1+n-1)}$ of admissible symbolic words of length
$n$ consists of approximately $2^{nh}$ symbolic words, each approximately
of measure $2^{-nh}$, whereas each of the remaining symbolic words has
negligible measure.  The choice of $n_1$ is irrelevant because the
source is stationary.  Formally, a source has the asymptotic equipartition
property if and only if for any $\epsilon>0$ there is a positive integer
$n_0(\epsilon)$ such that, for $n>n_0(\epsilon)$ and arbitrary $n_1$,
the set $\Sigma_{\cal E}^{(n_1,n_1+n-1)}$ of admissible symbolic words
of length $n$ decomposes into two sets $\Pi$ and $T$ satisfying
\begin{equation}
\sum_{\tilde\omega\in\Pi} \mu(\tilde\omega) < \epsilon
\end{equation}
and
\begin{equation}
2^{-n(h+\epsilon)} < \mu(\tilde\omega) < 2^{-n(h-\epsilon)}
\mbox{~~for all $\tilde\omega\in T$.}
\end{equation}

\section{Perturbed chaotic maps}  \label{secper}

Let $(M,\mu,f)$ be a discrete abstract dynamical system that is
derived from a Hamiltonian phase-space flow as described at the end of
Sec.~\ref{sechyp}.  This means, in particular, that the measure $\mu$
is the standard phase-space measure, in units such that the accessible
volume of phase space is unity.  At the $n$th step the effect of the
unperturbed system dynamics is to change the phase-space density
from the density $\rho(x,n-1)$ that emerges from the $(n-1)$th step
to  a new density
\begin{equation}
\rho'(x,n)=\rho(f^{-1}x,n-1)\;.
\end{equation}

We model a measure-preserving stochastic perturbation by alternating
unperturbed time steps with application of measure-preserving
{\it perturbation maps}.  More precisely, we do the following.  We
have available a collection of measure-preserving perturbation maps.
At the $n$th step we select randomly a particular perturbation map
$\xi:M\rightarrow M$ from this collection and apply it to the density
$\rho'(x,n)$ that is produced by the unperturbed time step.  This
yields a new density
\begin{equation}
\rho(x,n)=\rho'(\xi^{-1}x,n)=\rho\!\left(f^{-1}(\xi^{-1}x),n-1\right)\;,
\end{equation}
which depends on the map $\xi$ and which is the input to the next step.

We characterize the perturbation maps in terms of two quantities:
(i)~the ``strength'' of the perturbation, which is roughly the size
of the phase-space displacements produced by the maps, and (ii)~the
``correlation cells,'' which are roughly the phase-space regions
over which the displacements produced by the maps remain correlated.
We pause here to give a more precise general definition of perturbation
strength, because it highlights an essential feature of chaotic dynamics.
We defer defining the concept of correlation cells precisely till it
emerges naturally in the context of the symbolic dynamics of perturbed
chaotic maps.  We return to both concepts in Sec.~\ref{discussion},
where they are used to develop a heuristic picture of hypersensitivity
to perturbation.

To characterize the ``strength'' of a perturbation, we let
$\delta(x_1,x_2)$ denote the Euclidean distance between the two points
$x_1,x_2\in M$ relative to some fixed set of canonical co\"ordinates.
An {\it $\epsilon$-perturbation map\/} is a perturbation map
$\xi$ for which $\delta(\xi x,x)<\epsilon$ for all $x\in M$.  An
$\epsilon$-perturbation map describes a perturbation whose strength
is smaller than the scale set by $\epsilon$.

Now suppose that the initial density $\rho(x,\mbox{$n=0$})$ is well
behaved in the sense that there is a scale on which $\rho(x,\mbox{$n=0$})$
varies little; i.e., there is an $\epsilon_0>0$ such that
$\rho(x_1,\mbox{$n=0$})\simeq\rho(x_2 ,\mbox{$n=0$})$ for any pair of
points $x_1,x_2\in M$ with $\delta(x_1,x_2)<\epsilon_0$.  Then, for any
integer $n>0$, there is an $\epsilon>0$ such that $\rho(x,n)$ varies
little on the scale of $\epsilon$.  We say that the system is
{\it effectively shielded against perturbations\/} at the $n$th step
if there is an $\epsilon>0$ such that the perturbation is described
by $\epsilon$-perturbation maps and the density $\rho(x,n)$ varies
little on the scale of $\epsilon$.

One of the defining properties of chaotic dynamics is that the scale
$\epsilon$ on which the density varies little decreases exponentially
with the number of time steps $n$.  This entails that chaotic systems
cannot be effectively shielded against perturbations, except for a small
number of time steps.  We use this fact below as the starting point for
developing an essentially universal description of perturbed chaotic
dynamics.  Regular, or integrable systems have no exponential relationship
between $\epsilon$ and $n$ and thus cannot be fitted within the analysis of
this section.  We thus defer discussion of regular systems until we
have developed a heuristic picture of hypersensitivity to perturbation
in Sec.~\ref{discussion}.

We now proceed to show that all $K$ systems for which there is a
generating partition exhibit hypersensitivity to perturbation.  This
includes all $K$ systems that have a Markov partition \cite{Alekseev1981}.
Assume that the discrete abstract dynamical system $(M,\mu,f)$ has a finite
KS entropy $h=h_\mu(f)>0$, and let ${\cal E}=\{E_1,\ldots E_m\}$ be a
generating partition of $M$.  As explained in Sec.~\ref{secsym}, $f$
can be represented by a shift map $\sigma$ on the set of admissible
symbolic sequences $\Sigma_{\cal E}$, each admissible symbolic sequence
corresponding to a single point in $M$. In the following, we identify
symbolic sequences with the corresponding points and symbolic words
with the corresponding subsets of $M$, writing, e.g., ``the symbolic
word $\omega_{n_1}\cdots\omega_{n_2}$'' when we really mean the set of
points corresponding to the symbolic word $\omega_{n_1}\cdots\omega_{n_2}$.
The set of admissible symbolic sequences has the asymptotic equipartition
property; i.e., for $n\gg1$, $M$ is partitioned by the admissible
symbolic words $\omega_1\cdots\omega_n$ in such a way that there are
approximately $2^{nh}$ symbolic words each approximately of measure
$2^{-nh}$, whereas each of the remaining symbolic words has negligible
measure.

Let us first look at the unperturbed evolution of a simple initial state
on $M$.  We assume that the initial density $\rho(x,\mbox{$n=0$})$ is
constant on the set of points corresponding to the symbolic word
\begin{equation}
\Omega=\omega_{n_0+1}\cdots\omega_{n_0+q}\;,\;\;\mbox{$q\gg1$,}
\label{initpattern}
\end{equation}
and zero elsewhere.  Here $\omega_{n_0+1}\cdots\omega_{n_0+q}$ is one
of the symbolic words that has measure
\begin{equation}
\mu_0\equiv\mu(\omega_{n_0+1}\cdots\omega_{n_0+q})\simeq 2^{-qh}\;.
\end{equation}
In the following, we refer to a subset of $M$ on which the density is
constant as a {\it pattern}.

We choose the (arbitrary) zero of the entropy such that the entropy of
a uniform density constant on the entire set $M$ vanishes.  This is a
natural choice because it corresponds to choosing units such that
$\mu(x)$ is the measure in the Gibbs entropy~(\ref{Gibbsent}).  The
entropy of the initial density is thus
\begin{equation}
H_0=\log_2\mu_0\simeq\log_2(2^{-qh})=-qh \;.
\end{equation}
The condition $q\gg1$ means that the initial entropy $H_0$ is much
smaller than the negative of the KS entropy of the map, $-h$.

Applying the shift map $\sigma$ for $n$ steps leads to a uniform density
on $\Omega'=\sigma^n\Omega=\omega'_{n_0+1-n}\cdots\omega'_{n_0+q-n}$ where
$\omega'_k=\omega_{k+n}$. The entropy of the shifted pattern remains
unchanged.  As was stressed in Sec.~\ref{secstat}, the entropy does not
change under unperturbed Hamiltonian evolution.  Moreover, the method of
symbolic dynamics makes it utterly obvious that {\it no additional
information beyond the initial pattern and the number of steps $n$ is
needed to give a complete description of the evolved pattern}.  As was
pointed out in Sec.~\ref{sechyp}, the evolved unperturbed density, though
highly structured when viewed in phase space, is not complex in the
algorithmic sense.

We now turn to perturbed evolution.  At each step, instead of applying
just the map $f$, we now apply first $f$ and then a measure-preserving
map $\xi$ selected randomly from our collection of maps.  We make two
major assumptions about the perturbation maps $\xi$ and their probabilities,
the first assumption having to do with the perturbation strength and
the second with the perturbation correlation cells.  The first
assumption is that below some scale on phase space, a single application
of the perturbation randomizes the pattern completely.  In symbolic
language this scale is characterized by some negative integer $-n_p$,
and our assumption can be written as
\begin{eqnarray}
\mbox{Prob}\Bigl((\xi\omega)_k=\omega'_k,\;k&=&n,\ldots,-n_p\Bigr) =
   \mu(\omega'_n\cdots\omega'_{-n_p}|\,\omega_{-n_p+1}\omega_{-n_p+2}\cdots)
   \label{eqran}\\
&\mbox{}&\mbox{for all $n\leq-n_p$, $\omega\in\Sigma_{\cal E}$, and
$\omega'_n\cdots\omega'_{-n_p} \in \Sigma_{{\cal E}}^{(n,-n_p)}$}\;,
\nonumber
\end{eqnarray}
where Prob stands for probability with respect to the random selection of
the perturbation map.  The integer $n_p$  is a measure of the strength
of the perturbation, large $n_p$ meaning a weak perturbation.  Another
way of describing this assumption is the following: take a point on
phase space---i.e., a symbolic sequence $\omega$---and perturb it
to get a new point $\xi\omega$; Eq.~(\ref{eqran}) means that it is
unpredictable, relative to the random selection of perturbation map $\xi$,
in which partition element $E_i$ the $n_p$th backward iterate and all
further backward iterates of the perturbed point $\xi\omega$ fall.

Our second major assumption concerns the perturbation correlation cells.
We assume that the perturbation maps $\xi$ are not completely arbitrary,
but that a particular map displaces neighboring points in a similar way
and that, averaged over the random selection of perturbation maps, the
displacements become uncorrelated for points sufficiently far away from
each other.  We model this behavior by assuming that the space $M$ is
partitioned into {\it perturbation cells\/}
$\omega_{-n_p+s+1}\cdots\omega_{-n_p+s+r}$, where $r\gg1$ and $s\geq0$
are integers, such that, first, the perturbations are uncorrelated for
points in different perturbation cells and, second, knowing how a
typical point in a perturbation cell is perturbed determines how all
points in that cell are perturbed.  These perturbation cells are a
precise realization of the notion of correlation cells.

In addition to our two major assumptions, we make several simplifying
assumptions or approximations about the perturbation maps.  These
simplifying assumptions always tend to reduce the information
$\Delta I_{\rm min}$ required to reduce the entropy increase to the
tolerable amount $\Delta H_{\rm tol}$.  Since we want to prove that
$\Delta I_{\rm min}$ is large, such simplifying assumptions do not
limit the validity of our results.  As our first simplifying assumption,
we ignore all features of the perturbation maps beyond what is needed
to satisfy Eq.~(\ref{eqran}); i.e., we choose perturbation maps $\xi$
that satisfy
\begin{equation}
(\xi\omega)_n=\omega_n\;\;
\mbox{for all $n>-n_p$ and $\omega\in\Sigma_{\cal E}$,}
\label{pmaps}
\end{equation}
in addition to Eq.~(\ref{eqran}).  This assumption means that the
perturbation maps have no effect at all on scales larger than the scale
set by $n_p$.  Allowing the perturbation maps to act on scales larger
than that set by $n_p$ would lead to more distinguishable perturbed
patterns---and thus to higher $\Delta I_{\rm min}$---which would have
to be tracked to keep the entropy increase to some tolerable amount.

Since it is impossible to shield a chaotic system against perturbations
in the sense defined above, we are justified in choosing the zero of time
($n=0$) such that the perturbation becomes effective at the first time
step ($n=1$).  This amounts to choosing the initial symbolic
word~(\ref{initpattern}) so that $n_0=-n_p$, where $n_p$ is the integer
that characterizes the strength of the perturbation.  This initial symbolic
word, which defines the pattern on which the initial density
$\rho(x,\mbox{$n=0$})$ is nonzero, can thus be written as
\begin{equation}
\Omega=|\;\omega_{-n_p+1}\cdots\omega_{-n_p+s}\;|\;\omega_{-n_p+s+1}\cdots
\omega_{-n_p+s+r}\;|\;\cdots\omega_{-n_p+q}\;.
\label{eqregions}
\end{equation}
Since the perturbation maps satisfy Eq.~(\ref{pmaps}), the perturbation
leaves the pattern of Eq.~(\ref{eqregions}) unchanged.  After one time
step, however, the leftmost symbol moves into the {\it perturbation region},
located to the left of the leftmost vertical bar in Eq.~(\ref{eqregions}),
where it is randomized by the perturbation according to Eq.~(\ref{eqran}).
The perturbation region is separated by $s$ letters from the {\it decision
region}, located between the middle and rightmost vertical bars in
Eq.~(\ref{eqregions}).  This decision region, $r$ letters wide, defines
the perturbation cells.  Since we assume that $r\gg1$, there are
approximately $2^{rh}$ {\it typical perturbation cells}, each of size
$\simeq2^{-rh}$, whereas the total size of the remaining perturbation
cells can be neglected.  Even though the assumption $q>r+s$ is implicit
in the way we write the initial word in Eq.~(\ref{eqregions}), this
assumption is not necessary for our analysis.

Focus attention now on the phase-space density $\rho(x,n)$ after $n$
time steps, where we assume that
\begin{equation}
q-s\ge n\gg\max\Bigl(1,q-s-r\Bigr)\;.
\label{nlimits}
\end{equation}
These assumptions assure us that the leftmost letter of the initial
word~(\ref{eqregions}) has moved deep into the perturbation region and
the rightmost letter has moved far to the left of the right boundary of
the decision region, but not more than one position beyond the left
boundary of the decision region.  After $n$ {\it unperturbed\/} steps
the initial pattern $\Omega$ given by Eq.~(\ref{eqregions}) evolves into
the pattern $\Omega'=\sigma^n\Omega$, which has the form
\begin{equation}
\Omega'=
\omega'_{-n_p-n+1}\cdots\omega'_{-n_p}\;|\;\omega'_{-n_p+1}\cdots
\omega'_{-n_p+s}\;|\;\omega'_{-n_p+s+1}\cdots\omega'_{-n_p-n+q}\;,
\label{eqregions-n}
\end{equation}
where $\omega'_k=\omega_{k+n}$.

Consider now what happens when the pattern of Eq.~(\ref{eqregions-n}) is
perturbed.  According to Eq.~(\ref{eqran}), all $n$ letters in the
perturbation region [to the left of the leftmost vertical bar in
Eq.~(\ref{eqregions-n})] are randomized by the perturbation.  We can
therefore ignore the effect of perturbations applied at previous steps.
The density that arises from averaging over the perturbation is made up
of all the patterns that come from randomizing the letters in the
perturbation region.  As a consequence of the asymptotic equipartition
property and assumption~(\ref{eqran}), there are approximately $2^{nh}$
such patterns, all of which have approximately the same probability and
all of which have approximately the same measure as the unperturbed
pattern~(\ref{eqregions-n}).  Thus averaging over the perturbation leads
to an entropy increase
\begin{equation}
\Delta H_{\cal S}\simeq \log_2(2^{nh})=nh \;.
\label{HSnh}
\end{equation}

We now turn to estimating the minimum information $\Delta I_{\rm min}$
about the perturbation needed to limit the entropy increase to a tolerable
value $\Delta H_{\rm tol}$.  Consider again the word~(\ref{eqregions-n})
that describes the unperturbed pattern after $n$ steps.  Due to the
asymptotic equipartition property, the $n-(q-s-r)\gg1$ unspecified
letters at the right side of the decision region correspond to the
pattern's extending over
\begin{equation}
{\cal R}_n\equiv 2^{[n-(q-s-r)]h}\gg1
\label{rsubn}
\end{equation}
typical perturbation cells.  This exponential increase in the number of
typical perturbation cells occupied by the pattern continues only until
all the typical perturbation cells are occupied, i.e., until
${\cal R}_n=2^{rh}$ or $n=q-s$.  The occupied perturbation cells partition
the unperturbed pattern into ${\cal R}_n$ {\it sub-patterns\/} of the form
\begin{equation}
\omega'_{-n_p-n+1}\cdots\omega'_{-n_p}\;|\;\omega'_{-n_p+1}\cdots
\omega'_{-n_p+s}\;|\;\omega'_{-n_p+s+1}\cdots\omega'_{-n_p-n+q}
\hat\omega_{-n_p-n+q+1}\cdots\hat\omega_{-n_p+s+r}\;| \;,
\label{eqregions2}
\end{equation}
where the $n-(q-s-r)$ letters $\hat\omega_i$ determine an occupied
perturbation cell.

These sub-patterns, all of approximately the same size, are perturbed
independently.  We describe the {\it perturbed\/} sub-pattern in
each perturbation cell by a symbolic word
\begin{equation}
\tilde\omega_{-n_p-n+1}\cdots\tilde\omega_{-n_p}\;|\;\omega'_{-n_p+1}\cdots
\omega'_{-n_p+s}\;|\;\omega'_{-n_p+s+1}\cdots\omega'_{-n_p-n+q}
\hat\omega_{-n_p-n+q+1}\cdots\hat\omega_{-n_p+s+r}\;| \;,
\label{eqregions3}
\end{equation}
where the letters $\tilde\omega_i$ are chosen at random according to
Eq.~(\ref{eqran}).  Again invoking the asymptotic equipartition property,
we can say that in each of the ${\cal R}_n$ occupied perturbation cells,
there are
\begin{equation}
D\equiv2^{nh}\gg1
\label{bigD}
\end{equation}
typical perturbed words, or typical perturbed sub-patterns, of the
form~(\ref{eqregions3}), all having approximately the same probability
and all having approximately the same measure as the unperturbed
sub-pattern~(\ref{eqregions2}).

These considerations give a total of $D^{{\cal R}_n}$ typical perturbed
patterns, all produced with approximately the same probability by the
perturbation and all having approximately the same entropy as the
unperturbed pattern~(\ref{eqregions-n}).  The information needed to
specify a particular perturbed pattern---and thus the information
needed to keep the tolerable entropy increase essentially to zero---is
given by
\begin{equation}
\Delta I_{\rm min} \simeq {\cal R}_n\log_2D
\simeq 2^{[n-(q-s-r)]h}\Delta H_{\cal S}\;\;\;
\mbox{for $\Delta H_{\rm tol}\simeq0$.}
\label{logpatt}
\end{equation}
It should be emphasized that the exponential increase of this
$\Delta I_{\rm min}$ continues only until all the typical perturbation
cells are occupied, i.e., until $n=q-s$; for $n>q-s$ the information
continues to increase, but the form of the increase is more difficult
to determine.

What is going on here has a simple interpretation.  Within each
perturbation cell, the perturbed sub-patterns have essentially no
overlap.  The overall perturbed patterns, however, can have considerable
overlap, since two perturbed patterns are different even if they
differ in only a single perturbation cell.  The entropy increase
$\Delta H_{\cal S}\simeq nh$ that comes from averaging over the
perturbation [Eq.~(\ref{HSnh})] is the logarithm of the number $D$ of
{\it non-overlapping\/} patterns that are required to make up the
average density.  The number of non-overlapping patterns is the same
as the number of perturbed sub-patterns in each perturbation cell, and
hence $\Delta H_{\cal S}\simeq nh$ is also the information required to
specify a particular sub-pattern within a perturbation cell.  To specify
a particular overall pattern, however, one must say which perturbed
sub-pattern is realized in each of the ${\cal R}_n$ occupied perturbation
cells; this requires giving $\Delta H_{\cal S}\simeq nh$ bits per occupied
perturbation cell, for a total amount of information
$\Delta I_{\rm min}\simeq{\cal R}_n\Delta H_{\cal S}$ [Eq.~(\ref{logpatt})].
The information $\Delta I_{\rm min}$ is much bigger than the average
entropy increase $\Delta H_{\cal S}$ because the information counts
overlapping patterns, whereas the entropy does not.

Now suppose that one allows a nonzero tolerable entropy increase
$\Delta H_{\rm tol}$.  This means that one does not have to specify
exactly which of the $D^{{\cal R}_n}$ perturbed patterns is realized.
Instead, one can group the typical perturbed patterns and specify only
to which group the perturbed pattern belongs.  Suppose the typical
patterns are grouped into $N$ groups, which are labeled by an integer
$b=1,\ldots,N$.  In analogy to Sec.~\ref{sechyp}, we denote by $N_b$
the number of patterns in the $b$th group ($\sum_{b=1}^N N_b=D^{{\cal R}_n}$),
by $\rho_b(x)$ the probability density one obtains by averaging over
all the patterns in the $b$th group, by $\Delta H_b$ the corresponding
conditional entropy increase, and by $\overline{\Delta H}=
\sum p_b\Delta H_b$ the average conditional entropy increase.  Since
all the patterns are approximately equi-probable, the probability of
obtaining the measurement record $b$, which specifies that the perturbed
pattern is in the $b$th group, is $p_b=N_bD^{-{\cal R}_n}$.

To obtain $\Delta I_{\rm min}$ for a given $\Delta H_{\rm tol}$, one
would have to find a grouping of the patterns that is optimal in the
sense of minimizing $\Delta I_{\rm min}$ under the condition that
$\overline{\Delta H}\leq\Delta H_{\rm tol}$.  Since we do not know how
to find an optimal grouping, we construct a nearly optimal grouping as
follows.  We start with a particular pattern, or {\it fiducial pattern},
and form our first group out of all the patterns that differ in at most
$d$ perturbation cells from the fiducial pattern.  Such a group we call
a $d$-group.  The grouping into $d$-groups is motivated by the fact
that the entropy increase $\overline{\Delta H}$ is minimal for groups
of patterns that differ in the smallest number of perturbation cells
[see Eq.~(\ref{Hdapprox1}) below].  There being
\begin{equation}
g_k={{\cal R}_n\choose k}(D-1)^k
\end{equation}
patterns that differ in exactly $k$ cells from an arbitrary fiducial
pattern, the number of patterns differing in at most $d$ cells from an
arbitrary fiducial pattern and therefore the size of a $d$-group is
\begin{equation}
G_d= \sum_{k=0}^d g_k = \sum_{k=0}^d {{\cal R}_n\choose k}(D-1)^k \;.
\end{equation}

A particularly simple way to proceed would be to pick a second fiducial
pattern from among the patterns not in the first group, forming a
second $d$-group about this second pattern, and then to continue to
form $d$-groups until all patterns were grouped.  Unfortunately, this
strategy fails because if we proceed in this way, some groups overlap.
The problem of finding a grouping into non-overlapping $d$-groups is
equivalent to the problem of finding a perfect error-correcting code in
information theory \cite{Welsh1988} and generally has no solution.  In
the following, we nevertheless assume that the $D^{{\cal R}_n}$ patterns
are perfectly grouped into a number $N=D^{{\cal R}_n}/G_d$ of $d$-groups.
We can make this simplifying assumption because it lowers our estimate of
$\Delta I_{\rm min}$.

We now turn to the computation of the entropy increase $\Delta H_d$ for
a $d$-group, i.e., a group consisting of a fiducial pattern and all
the patterns differing in at most $d$ perturbation cells from the
fiducial pattern.   The average density $\rho_d(x)$ for a $d$-group
is the average of the densities for the $G_d$ patterns in the group,
all patterns contributing with the same probability $1/G_d$.
Alternatively, we can break each contributing pattern into its ${\cal R}_n$
sub-patterns---i.e., symbolic words of the form~(\ref{eqregions3})---and
view $\rho_d(x)$ as being made up of contributions from the
$D{\cal R}_n$ sub-patterns, all of which have approximately the same
measure $\mu_0/{\cal R}_n$.

We distinguish two types of sub-patterns, namely the ${\cal R}_n$
sub-patterns belonging to the fiducial pattern and the other
$(D-1){\cal R}_n$ sub-patterns.  The average density $\rho_d(x)$ is
uniform on each sub-pattern.  We denote its value on sub-patterns
belonging to the fiducial sub-pattern by $\rho_{df}$ and its value on
the other sub-patterns by $\rho_{do}$.  For a sub-pattern belonging
to the fiducial pattern, the probability obtained by integrating
$\rho_d$ over the sub-pattern is
\begin{equation}
\int d\mu(x)\,\rho_d=\rho_{df}{\mu_0\over{\cal R}_n}={p_f\over{\cal R}_n}\;,
\end{equation}
where $p_f$ is the probability obtained by integrating $\rho_d$ over
the entire fiducial pattern.  Similarly, for any of the other sub-patterns,
the probability obtained by integrating $\rho_d$ over the sub-pattern is
\begin{equation}
\int d\mu(x)\,\rho_d=\rho_{do}{\mu_0\over{\cal R}_n}=
{p_o\over (D-1){\cal R}_n}\;,
\end{equation}
where $p_o=1-p_f$ is the probability obtained by integrating $\rho_d$ over
all the sub-patterns outside the fiducial pattern.

The entropy increase of a $d$-group can now be written as
\begin{eqnarray}
\Delta H_d&=&-\int d\mu(x)\,\rho_d(x)\log_2[\rho_d(x)]\;-H_0\nonumber\\
&=&-{\cal R}_n{p_f\over{\cal R}_n}\log_2\!\left({p_f\over\mu_0}\right)
-(D-1){\cal R}_n{p_o\over(D-1){\cal R}_n}
\log_2\!\left({p_o\over\mu_0(D-1)}\right)
-\log_2\mu_0\nonumber\\
&=&-p_f\log_2 p_f-p_o\log_2p_o+p_o\log_2(D-1)\;.
\end{eqnarray}
To evaluate $\Delta H_d$, we must find the integrated probabilities
$p_o$ and $p_f$.  Each pattern that differs in exactly $k$ cells from the
fiducial pattern contributes the amount $k/{\cal R}_nG_d$ to $p_o$ and
the amount $({\cal R}_n-k)/{\cal R}_nG_d$ to $p_f$.  It follows that
\begin{equation}
p_o=\sum_{k=0}^d {k\over{\cal R}_nG_d}g_k=
{1\over{\cal R}_nG_d}\, \sum_{k=0}^d{{\cal R}_n\choose k} (D-1)^k \,k
\end{equation}
and
\begin{equation}
p_f=\sum_{k=0}^d{{\cal R}_n-k\over{\cal R}_nG_d}g_k=
{1\over{\cal R}_nG_d}\, \sum_{k=0}^d{{\cal R}_n\choose k}
(D-1)^k \,({\cal R}_n-k)\;.
\end{equation}
Notice that $p_o={\cal R}_n^{-1}\partial\ln G_d/\partial\ln(D-1)$
and that when $d={\cal R}_n$, we have $G_{{\cal R}_n}=D^{{\cal R}_n}$,
$p_o=1-1/D$, $p_f=1/D$, and thus $\Delta H_{{\cal R}_n}=\log_2D=
\Delta H_{{\cal S}}$.

Under the assumption of perfect grouping into $N=D^{{\cal R}_n}/G_d$
$d$-groups, the average entropy of the $d$-groups is
$\overline{\Delta H}=\Delta H_d$, and the information to specify a
particular $d$-group is
\begin{equation}
\Delta I_d \simeq \log_2 N
\simeq{\cal R}_n\log_2 D-\log_2G_d={\cal R}_n\Delta H_{\cal S}-\log_2 G_d \;.
\label{Id}
\end{equation}
Under our further simplifying assumption that optimal grouping is well
approximated by perfect grouping into $d$-groups, we can approximate
the minimum information $\Delta I_{\rm min}$ required to keep the entropy
increase to a tolerable amount $\Delta H_{\rm tol}$ by
\begin{equation}
\Delta I_{\rm min}\simeq\Delta I_d\;\;
\mbox{for $\Delta H_{\rm tol}\simeq\Delta H_d\;.$}
\label{Iapprox}
\end{equation}

At this point we could plot $\Delta I_{\rm min}$ as a function of
$\Delta H_{\rm tol}$ by using the common dependence on $d$.  Given
the assumptions~(\ref{rsubn}) and (\ref{bigD}) that ${\cal R}_n$ and
$D$ are large, however, we can introduce further approximations that
allow us to write an explicit expression for $\Delta I_{\rm min}$ as
a function of $\Delta H_{\rm tol}$, valid over nearly the entire range
of $\Delta H_{\rm tol}$.  The key to these approximations is that $g_k$
increases exponentially for $k\ll k_c\equiv({\cal R}_n+1)(1-1/D)$.
This means that each of the sums for $G_d$, $p_o$, and $p_f$ can be
approximated by its largest term ($k=d$), provided
${\cal R}_n-d\gg{\cal R}_n-k_c\simeq{\cal R}_n/D-1$.   The resulting
approximations
are
\begin{equation}
G_d\simeq {{\cal R}_n\choose d}(D-1)^d, \;\;\;\;
p_o\simeq{d\over{\cal R}_n}, \;\;\;\;
p_f\simeq{{\cal R}_n-d\over{\cal R}_n} \;.
\label{keyapprox}
\end{equation}
In this approximation the entropy increase of a $d$-group is
\begin{equation}
\Delta H_d\simeq
-{{\cal R}_n-d\over{\cal R}_n}\log_2{{\cal R}_n-d\over{\cal R}_n}
-{d\over{\cal R}_n}\log_2{d\over{\cal R}_n}
+{d\over{\cal R}_n}\log_2(D-1)\;.
\label{Hdapprox1}
\end{equation}

Using the same approximation and applying Stirling's formula, one finds
that
\begin{eqnarray}
\log_2G_d&\simeq&\log_2\!{{\cal R}_n\choose d}+d\log_2(D-1)\nonumber\\
&\simeq&{\cal R}_n\!\left(
-{{\cal R}_n-d\over{\cal R}_n}\log_2{{\cal R}_n-d\over{\cal R}_n}
-{d\over{\cal R}_n}\log_2{d\over{\cal R}_n}
+{d\over{\cal R}_n}\log_2(D-1)\right)\nonumber\\
&\simeq&{\cal R}_n\Delta H_d\;.
\label{Hdapprox}
\end{eqnarray}
Combining Eqs.~(\ref{Id}), (\ref{Iapprox}), and (\ref{Hdapprox}) yields
\begin{equation}
\Delta I_{\rm min}\simeq\Delta I_d
\simeq{\cal R}_n(\Delta H_{\cal S}-\Delta H_d)
\simeq{\cal R}_n(\Delta H_{\cal S}-\Delta H_{\rm tol})\;.
\label{keyresult}
\end{equation}
This expression, the key result of this paper, shows that to reduce
the entropy of a perturbed chaotic map by an amount
$\Delta H_{\cal S}-\Delta H_{\rm tol}$, one must acquire an amount of
information $\Delta I_{\rm min}$ about the perturbation which is much
larger than the contemplated entropy reduction.  Indeed, the ratio of
information to entropy reduction grows exponentially as
${\cal R}_n=2^{[n-(q-s-r)]h}$ with the number of time steps, the exponential
rate of growth being determined by the KS entropy $h$ of the map.  This
is what we mean by exponential hypersensitivity to perturbation.

We should investigate the validity of the approximations that lead to
our key result~(\ref{keyresult}).  This result agrees with what we have
already derived in Eq.~(\ref{logpatt}) for $\Delta H_{\rm tol}\simeq0$.
Thus we are mainly interested in knowing where the approximations fail
as $\Delta H_{\rm tol}$ approaches $\Delta H_{\cal S}$.  A more careful
analysis, which keeps track of the errors introduced by the
approximation~(\ref{keyapprox}) and by the use of Stirling's formula
in Eq.~(\ref{Hdapprox}), indicates that we must consider separately
two cases: (i)~${\cal R}_n\agt D$ ($r+s\agt q$), i.e., there are more
occupied perturbation cells than there are perturbed sub-patterns per
cell; (ii)~${\cal R}_n\alt D$ ($r+s\alt q$), i.e., there are fewer
occupied perturbation cells than there are perturbed sub-patterns per
cell.  In case~(i), Eq.~(\ref{keyresult}) is valid as long as
${\cal R}_n-d\gg{\cal R}_n/D\agt1$, which translates to
\begin{equation}
\mbox{${\cal R}_n\agt D$:}\;\;\;
\Delta H_{\cal S}-\Delta H_{\rm tol}\gg{1\over D}
\;\Longleftrightarrow\;
\Delta I_{\rm min}\gg{{\cal R}_n\over D}\agt1\;;
\label{rngreaterD}
\end{equation}
in case~(ii), Eq.~(\ref{keyresult}) is valid as long as
${\cal R}_n-d\gg\ln(eD/{\cal R}_n)\agt1\agt{\cal R}_n/D$, which
translates to
\begin{equation}
\mbox{${\cal R}_n\alt D$:}\;\;\;
\Delta H_{\cal S}-\Delta H_{\rm tol}\gg
{1\over{\cal R}_n}\Biggl(\ln\!\left({eD\over{\cal R}_n}\right)\Biggr)^2
\;\Longleftrightarrow\;
\Delta I_{\rm min}\gg\Biggl(\ln\!\left({eD\over{\cal R}_n}\right)\Biggr)^2
\agt1\;.
\label{rnlessD}
\end{equation}
These restrictions arise because of approximations made in evaluating
$\Delta I_d$ and $\Delta H_d$.

There is a separate question of whether perfect $d$-grouping is a good
approximation to optimal grouping.  The restrictions contained in
Eqs.~(\ref{rngreaterD}) and (\ref{rnlessD}) are probably not the most
important restrictions on the validity of our key result, because the
very idea of perfect $d$-grouping as an approximation to optimal grouping is
suspect when $\Delta I_{\rm min}$ is as small as a few bits.  Our hesitancy
in defining exponential hypersensitivity to perturbation, where we
require the information-to-entropy ratio~(\ref{eqhyp}) to grow
exponentially for ``almost all'' values of $\Delta H_{\rm tol}$, can be
traced to this inability to approximate the optimal grouping when
$\Delta H_{\rm tol}$ is very close to $\Delta H_{{\cal S}}$.  We are
left uncertain about the precise behavior of $\Delta I_{\rm min}$ when
$\Delta H_{\rm tol}$ is very close to $\Delta H_{{\cal S}}$.

We can interpret our key result by hearkening back to the interpretation
given to Eq.~(\ref{logpatt}).  We first need to describe what it means to
to specify the phase-space density at a level of resolution defined by
a tolerable entropy increase $\Delta H_{\rm tol}$.  To do so, imagine that
the sub-patterns within each occupied perturbation cell are aggregated
into groups, which we call {\it coarse-grained sub-patterns}, each group
consisting of $2^{\Delta H_{\rm tol}}$ sub-patterns so that there are
\begin{equation}
{\cal D}=D/2^{\Delta H_{\rm tol}}
\simeq 2^{\Delta H_{\cal S}-\Delta H_{\rm tol}}
\end{equation}
coarse-grained sub-patterns in each occupied perturbation cell.  A
{\it coarse-grained pattern\/} consists of coarse-grained sub-patterns,
one for each of the ${\cal R}_n$ occupied perturbation cells.  Since a
coarse-grained pattern has a measure that is approximately
$2^{\Delta H_{\rm tol}}$ times as big as a pattern, a coarse-grained
pattern represents an entropy increase
\begin{equation}
\log_2(2^{\Delta H_{\rm tol}})=\Delta H_{\rm tol}\;.
\end{equation}
Thus, specifying the system state at a level of resolution set by
$\Delta H_{\rm tol}$ amounts to specifying a particular coarse-grained
pattern.

The further entropy increase that results from averaging over the
coarse-grained patterns is given approximately by
\begin{equation}
\log_2{\cal D}\simeq\Delta H_{\cal S}-\Delta H_{\rm tol}\;.
\label{HSminusHtol}
\end{equation}
This entropy increase is the logarithm of the number of
{\it non-overlapping\/} coarse-grained patterns that are required to
make up the density that comes from averaging over the perturbation.
This number of non-overlapping coarse-grained patterns is the same
as the number of coarse-grained sub-patterns in each perturbation cell,
and hence the entropy increase~(\ref{HSminusHtol}) is also the
information required to specify a particular coarse-grained sub-pattern
within a perturbation cell.  There being ${\cal R}_n$ perturbation cells,
the information needed to specify an entire coarse-grained pattern
becomes
\begin{equation}
\Delta I_{\rm min}\simeq{\cal R}_n(\Delta H_{\cal S}-\Delta H_{\rm tol})
={\cal R}_n\log_2{\cal D}\;,
\label{exphyp}
\end{equation}
an amount of information that corresponds to a total of
${\cal D}^{{\cal R}_n}$ coarse-grained patterns, all produced with
approximately the same probability by the perturbation.

The exponential hypersensitivity to perturbation that we have demonstrated
here for maps with positive KS entropy is an asymptotic property for
large times.  By spelling out precisely the character of the
$n\rightarrow\infty$ limit, we can see how exponential hypersensitivity
to perturbation provides an alternative definition of the KS entropy.
In discussing the limit, it is helpful to have in mind the
form~(\ref{eqregions}) of the initial symbolic word and the
form~(\ref{eqregions-n}) of the unperturbed symbolic word after $n$
time steps.  The assumptions~(\ref{nlimits}) indicate that as $n$ goes
to infinity, we should let $n-(q-r-s)$ go to infinity in the same
way as $n$---this allows the limit to explore the long-time exponential
growth of ${\cal R}_n$---while keeping $q-s-n\ge0$ constant---this
prevents the exponential growth of ${\cal R}_n$ from being halted at
the time when there is more than one sub-pattern per perturbation cell.
Thus an appropriate limit is to let $n$, $q$, and $r$ go to infinity,
while keeping $s$, $q-n\ge s$, and $r-n$ constant.  In thinking about
how this limit is mapped onto phase space, it is convenient also to
let $n_p$ go to infinity while keeping $-n_p+q$ constant; this keeps
the rightmost letter of the initial symbolic word in the same place
as we take the limit.  With this understanding of the limit, we can write
\begin{equation}
\lim_{n\rightarrow\infty}\Biggl({1\over n}
\log_2\!\left(
{\Delta I_{\rm min}\over\Delta H_{\cal S}-\Delta H_{\rm tol}}
\right)\Biggr)
=\lim_{n\rightarrow\infty}\!\left({\log_2{\cal R}_n\over n}\right)=h\;.
\end{equation}
In terms of phase space, this long-time limit means that the size
of the initial pattern, the size of a typical perturbation cell, and
the strength of the perturbation all go to zero at the same rate as $n$
goes to infinity.

\section{Discussion} \label{discussion}

The objective of this final section is to extract the important ideas
from the symbolic dynamics and to use them to develop a simple, heuristic
picture of hypersensitivity to perturbation.  Consider a classical
system whose dynamics unfolds on a $2F$-dimensional phase space, and
suppose that the system is perturbed by a stochastic Hamiltonian
whose effect can be described as diffusion on phase space.

Suppose first that the system is globally chaotic with KS entropy~$K$.
For such a system a phase-space density is stretched and folded by
the chaotic dynamics, developing exponentially fine structure as the
dynamics proceeds.  A simple picture is that the phase-space density
stretches exponentially in half the phase-space dimensions and contracts
exponentially in the other half of the dimensions.

The perturbation is characterized by a perturbation strength and by
correlation cells.  We can take the perturbation strength to be the
typical distance (e.g., Euclidean distance with respect to some fixed
set of canonical co\"ordinates) that a phase-space point diffuses under
the perturbation during an $e$-folding time, $F/K\ln2$, in a typical
contracting dimension.  The perturbation becomes effective, in the
sense described in Sec.~\ref{secper}, when the phase-space density has
roughly the same size in the contracting dimensions as the perturbation
strength.  Once the perturbation becomes effective, the effects of the
diffusive perturbation and of the further contraction roughly balance
one another, leaving the {\it average\/} phase-space density with a
constant size in the contracting dimensions.

The correlation cells are phase-space cells over which the effects of
the perturbation are well correlated and between which the effects of the
perturbation are essentially uncorrelated.  We assume that all the
correlation cells have approximately the same phase-space volume.  We
can get a rough idea of the effect of the perturbation by regarding
the correlation cells as receiving independent perturbations.  Moreover,
the diffusive effects of the perturbation during an $e$-folding time
$F/K\ln2$ are compressed exponentially during the next such
$e$-folding time; this means that once the perturbation becomes
effective, the main effects of the perturbation at a particular time
are due to the diffusion during the immediately preceding $e$-folding
time.

Since a chaotic system cannot be forever shielded from the effects of
the perturbation, we can choose the initial time $t=0$ to be the time at
which the perturbation is just becoming effective.  We suppose that
at $t=0$ the unperturbed density is spread over $2^{-Kt_0}$ correlation
cells, $t_0$ being the time when the unperturbed density occupies a
single correlation cell.  The essence of the KS entropy is that for
large times $t$ the unperturbed density spreads over
\begin{equation}
{\cal R}(t)\sim 2^{K(t-t_0)}
\end{equation}
correlation cells, in each of which it occupies roughly the same
phase-space volume.  The exponential increase of ${\cal R}(t)$ continues
until the unperturbed density is spread over essentially all the
correlation cells.  We can regard the unperturbed density as being
made up of {\it sub-densities}, one in each occupied correlation cell
and all having roughly the same phase-space volume.

After $t=0$, when the perturbation becomes effective, the {\it average\/}
density continues to spread exponentially in the expanding dimensions.
This spreading is not balanced, however, by contraction in the other
dimensions, so the phase-space volume occupied by the average density
grows as $2^{Kt}$, leading to an entropy increase
\begin{equation}
\Delta H_{\cal S}\sim\log_2(2^{Kt})=Kt\;.
\end{equation}
Just as the unperturbed density can be broken up into sub-densities,
so the average density can be broken up into {\it average sub-densities},
one in each occupied correlation cell.  Each average sub-density occupies
a phase-space volume that is $2^{Kt}$ times as big as the volume occupied
by an unperturbed sub-density.

The unperturbed density is embedded within the phase-space volume
occupied by the average density and itself occupies a volume that
is smaller by a factor of $2^{-Kt}$.  We can picture a {\it perturbed\/}
density crudely by imagining that in each occupied correlation cell
the unperturbed sub-density is moved rigidly to some new position within
the volume occupied by the average sub-density; the result is a
{\it perturbed sub-density}.  A {\it perturbed density\/} is made up
of perturbed sub-densities, one in each occupied correlation cell.  All
of the possible perturbed densities are produced by the perturbation
with roughly the same probability.

Suppose now that we wish to hold the entropy increase to a tolerable
amount $\Delta H_{\rm tol}$.  We must first describe what it means to
specify the phase-space density at a level of resolution set by a
tolerable entropy increase $\Delta H_{\rm tol}$.  An approximate
description can be obtained in the following way.  Take an occupied
correlation cell, and divide the volume occupied by the average
sub-density into $2^{\Delta H_{\cal S}-\Delta H_{\rm tol}}$ non-overlapping
volumes, all of the same size.  Aggregate all the perturbed sub-densities
that lie predominantly within a particular one of these non-overlapping
volumes to produce a {\it coarse-grained sub-density}.  There are
$2^{\Delta H_{\cal S}-\Delta H_{\rm tol}}$ coarse-grained sub-densities
within each occupied correlation cell, each having a phase-space volume
that is bigger than the volume occupied by a perturbed sub-density
by a factor of
\begin{equation}
{2^{Kt}\over2^{\Delta H_{\cal S}-\Delta H_{\rm tol}}}=
2^{\Delta H_{\rm tol}}\;.
\label{cgvolume}
\end{equation}
A {\it coarse-grained density\/} is made up by choosing a coarse-grained
sub-density in each occupied correlation cell.  A coarse-grained density
occupies a phase-space volume that is bigger than the volume occupied by
the unperturbed density by the factor $2^{\Delta H_{\rm tol}}$ of
Eq.~(\ref{cgvolume}) and hence represents an entropy increase
\begin{equation}
\log_2(2^{\Delta H_{\rm tol}})=\Delta H_{\rm tol}\;.
\end{equation}
Thus to specify the phase-space density at a level of resolution set
by $\Delta H_{\rm tol}$ means roughly to specify a coarse-grained
density.  The further entropy increase on averaging over the perturbation
is given by
\begin{equation}
\log_2(2^{\Delta H_{\cal S}-\Delta H_{\rm tol}})=
\Delta H_{\cal S}-\Delta H_{\rm tol}\;.
\label{furtherincrease}
\end{equation}

What about the information $\Delta I_{\rm min}$ required to hold the
entropy increase to $\Delta H_{\rm tol}$?  Since there are
$2^{\Delta H_{\cal S}-\Delta H_{\rm tol}}$ coarse-grained sub-densities
in an occupied correlation cell, each produced with roughly the same
probability by the perturbation, it takes approximately
$\Delta H_{\cal S}-\Delta H_{\rm tol}$ bits to specify a particular
coarse-grained sub-density.  To describe a coarse-grained density, one
must specify a coarse-grained sub-density in each of the ${\cal R}(t)$
occupied correlation cells.  Thus the information required to specify a
coarse-grained density---and, hence, the information required to hold
the entropy increase to $\Delta H_{\rm tol}$---is given by
\begin{equation}
\Delta I_{\rm min}\sim{\cal R}(t)(\Delta H_{\cal S}-\Delta H_{\rm tol})
\label{picsum}
\end{equation}
[cf.~Eq.~(\ref{exphyp})], corresponding to there being a total of
$(2^{\Delta H_{\cal S}-\Delta H_{\rm tol}})^{{\cal R}(t)}$ coarse-grained
densities.  The entropy increase~(\ref{furtherincrease}) comes from
counting the number of {\it non-overlapping\/} coarse-grained densities
that are required to fill the volume occupied by the average density,
that number being $2^{\Delta H_{\cal S}-\Delta H_{\rm tol}}$. In
contrast, the information $\Delta I_{\rm min}$ comes from counting
the exponentially greater number of ways of forming {\it overlapping\/}
coarse-grained densities by choosing one of the
$2^{\Delta H_{\cal S}-\Delta H_{\rm tol}}$ non-overlapping coarse-grained
sub-densities in each of the ${\cal R}(t)$ correlation cells.

The picture developed in this section, summarized neatly in
Eq.~(\ref{picsum}), requires that $\Delta H_{\rm tol}$ be big enough
that a coarse-grained sub-density is much larger than a perturbed
sub-density, so that we can talk meaningfully about the perturbed
sub-densities that lie predominantly {\it within\/} a coarse-grained
sub-density.  If $\Delta H_{\rm tol}$ becomes too small, Eq.~(\ref{picsum})
breaks down, and the information $\Delta I_{\rm min}$, rather than
reflecting a property of the chaotic dynamics as in Eq.~(\ref{picsum}),
becomes essentially a property of the perturbation, reflecting a
counting of the number of possible realizations of the perturbation.

The boundary between the two kinds of behavior of
$\Delta I_{\rm min}$ is set roughly by the number $F$ of contracting
phase-space dimensions.  When $\Delta H_{\rm tol}/F\agt1$, the
characteristic scale of a coarse-grained sub-density in the contracting
dimensions is a factor of
\begin{equation}
(2^{\Delta H_{\rm tol}})^{1/F}=2^{\Delta H_{\rm tol}/F}\agt2
\end{equation}
larger than the characteristic size of a perturbed sub-density in the
contracting dimensions.  In this regime the picture developed in this
section is at least approximately valid, because a coarse-grained
sub-density can accommodate several perturbed sub-densities in each
contracting dimension.  The information $\Delta I_{\rm min}$ becomes
a property of the system dynamics, rather than a property of the
perturbation, because it quantifies the effects of the perturbation
on scales as big as or bigger than the finest scale set by the system
dynamics.

In contrast, when $\Delta H_{\rm tol}/F\alt1$, we are required to
keep track of the phase-space density on a very fine scale in the
contracting dimensions, a scale smaller than the characteristic size
of a perturbed sub-density in the contracting dimensions.  Sub-densities
are considered to be distinct, even though they overlap substantially,
provided that they differ by more than this very fine scale in the
contracting dimensions.  The information $\Delta I_{\rm min}$ is
the logarithm of the number of realizations of the perturbation
which differ by more than this very fine scale in at least one
correlation cell.  The information becomes a property of the
perturbation because it reports on the effects of the perturbation
on scales finer than the finest scale set by the system dynamics---i.e.,
scales that are essentially irrelevant to the system dynamics.

We are now prepared to put in final form the exponential hypersensitivity
to perturbation of systems with a positive KS entropy:
\begin{equation}
{\Delta I_{\rm min}\over\Delta H_{\cal S}-\Delta H_{\rm tol}}\sim
{\cal R}(t)\sim 2^{K(t-t_0)}\;\;\;\mbox{for $\Delta H_{\rm tol}\agt F$.}
\label{picsum2}
\end{equation}
Once the chaotic dynamics renders the perturbation effective, this
exponential hypersensitivity to perturbation is essentially independent
of the form and strength of the perturbation.  Its essence is that within
each correlation cell there is a roughly even trade-off between entropy
reduction and information, but for the entire phase-space density the
trade-off is exponentially unfavorable because the density occupies an
exponentially increasing number of correlation cells, in each of which
it is perturbed independently.

As noted above, the behavior of $\Delta I_{\rm min}$ for
$\Delta H_{\rm tol}\alt F$ deviates from the universal behavior of
Eq.~(\ref{picsum2}) and tells one about the number of realizations
of the perturbation that produce densities that differ on scales finer
than the finest scale set by the system dynamics.  For a diffusive
perturbation of the sort contemplated in this section, $\Delta I_{\rm min}$
diverges as $\Delta H_{\rm tol}$ goes to zero, because a diffusive
perturbation has an infinite number of realizations on even the
tiniest scale.  If the diffusive perturbation is replaced by a
similar perturbation, but with a finite number of realizations, then
the growth of $\Delta I_{\rm min}$ is capped at the logarithm of the
number of realizations, corresponding to the finest scale on which the
perturbation acts.  The perturbation used in the symbolic-dynamics
analysis of perturbed chaotic maps in Sec.~\ref{secper} is of this
latter sort, with a finite number of realizations, the number being
$D^{{\cal R}_n}=(2^{nh})^{{\cal R}_n}$.  Indeed, the major simplifying
assumption about the perturbation in Sec.~\ref{secper} is that the
sub-patterns produced by the perturbation are all different on the finest
scale set by the system dynamics; i.e., there are no overlapping perturbed
sub-patterns. This means that the cap on
$\Delta I_{\rm min}$, which occurs at
$\Delta I_{\rm min}\simeq\log_2(D^{{\cal R}_n})={\cal R}_n nh$
[cf.~Eq.~(\ref{logpatt})], is such that the universal behavior of
Eq.~(\ref{exphyp}) extends right down to $\Delta H_{\rm tol}\simeq0$.

What about systems with regular, or integrable dynamics?  Though we
expect no universal behavior for regular systems, we can get an idea of
the possibilities from the heuristic description developed in this
section.  Hypersensitivity to perturbation requires, first, that the
phase-space density develop structure on the scale of the strength
of the perturbation, so that the perturbation becomes effective, and,
second, that after the perturbation becomes effective, the phase-space
density spread over many correlation cells.

For many regular systems there will be no hypersensitivity simply
because the phase-space density does not develop fine enough structure.
Regular dynamics can give rise to nonlinear shearing, however, in which
case the density can develop structure on the scale of the strength
of the perturbation and can spread over many correlation cells.  In this
situation, one expects the picture developed in this section to apply
at least approximately: to hold the entropy increase to
$\Delta H_{\rm tol}$ requires giving
$\Delta H_{\cal S}-\Delta H_{\rm tol}$ bits per occupied correlation
cell; $\Delta I_{\rm min}$ is related to $\Delta H_{\rm tol}$ by
Eq.~(\ref{picsum}), with ${\cal R}(t)$ being the number of correlation
cells occupied at time $t$.  Thus regular systems can display
hypersensitivity to perturbation if ${\cal R}(t)$ becomes large
(although this behavior could be eliminated by choosing correlation
cells that are aligned with the nonlinear shearing produced by the
system dynamics), but they cannot display {\it exponential\/}
hypersensitivity to perturbation because the growth of ${\cal R}(t)$
is slower than exponential.

A more direct way of stating this conclusion is to reiterate what we
have shown in this paper: Exponential hypersensitivity to perturbation
is equivalent to the spreading of phase-space densities over an
exponentially increasing number of phase-space cells; such exponential
spreading holds for chaotic, but not for regular systems and is
quantified by a positive value of the Kolmogorov-Sinai entropy.

\acknowledgments

The authors profited from discussions with H.~Barnum and R.~Menegus.

\end{document}